\documentclass[amsmath,amssymb,aps,letterpaper,prl,twocolumn,longbibliography]{revtex4-1}

\usepackage{graphicx}
\usepackage{color}
\usepackage{picture}
\usepackage{calc}
\usepackage{subfigure}

\usepackage{color}

\makeatletter \let\@keywords\@empty \let\@subject\@empty
\providecommand{\keywords}[1]{\gdef\@keywords{#1}}
\providecommand{\subject}[1]{\gdef\@subject{#1}}
\def\thetitle{\@title}
\def\theauthor{\@author}
\def\thesubject{\@subject}
\def\thedate{\@date}
\def\thekeywords{\@keywords}
\makeatother
\AtBeginDocument{%
\hypersetup{pdftitle={\thetitle}}%
\hypersetup{pdfauthor={\theauthor}}%
\hypersetup{pdfsubject={\thesubject}}%
\hypersetup{pdfkeywords={\thekeywords}}%
}

\usepackage[bookmarks=true,hyperfigures=true]{hyperref}
\usepackage{fixmath}
\usepackage{mathtools}

\providecommand{\href}[2]{#2}

\let\oldbfseries=\bfseries
\let\oldmdseries=\mdseries
\let\oldnormalfont=\normalfont
\renewcommand{\bfseries}{\oldbfseries\boldmath}
\renewcommand{\mdseries}{\oldmdseries\unboldmath}
\renewcommand{\normalfont}{\oldnormalfont\unboldmath}

\makeatletter
\newlength{\apb@width}
\newcommand{\autoparbox}[2][c]{\settowidth{\apb@width}{#2}\parbox[#1]{\apb@width}{#2}}

\makeatother

\RequirePackage{verbatim}

\makeatletter
\newwrite\bibinl@out
{\immediate\closeout\bibinl@out\@esphack}
\makeatother

\newcommand{\e}{\operatorname{e}}
\newcommand{\de}{\operatorname{d}\!}
\DeclareMathOperator{\tr}{tr}

\newcommand{\peps}{\varepsilon}

\newcommand{\gammaE}{\gamma_{\text{E}}}

\renewcommand{\digamma}{\Psi}

\newcommand{\calO}{\mathcal{O}}

\newcommand{\eqndot}{\, . }
\newcommand{\eqncom}{\, , }
\DeclareMathOperator{\cder}{D}
\newcommand{\scal}{\phi}
\newcommand{\scalc}{\scal^{\text{cl}}}
\newcommand{\scalq}{\tilde{\scal}}
\newcommand{\ferm}{\psi}
\newcommand{\aferm}{\bar{\ferm}}
\newcommand{\comm}[2]{[#1,#2]}

\newcommand{\YM}{{\mathrm{\scriptscriptstyle YM}}}

\begin{document}

\title{One-loop one-point functions in gauge-gravity dualities with defects}

\author{Isak Buhl-Mortensen}%
 \email{buhlmort@nbi.ku.dk}
   \author{Marius de Leeuw}%
    \email{deleeuwm@nbi.ku.dk}
       \author{Asger C.\ Ipsen}%
 \email{asgercro@nbi.ku.dk}
      \author{Charlotte Kristjansen}%
 \email{kristjan@nbi.ku.dk}
\author{Matthias Wilhelm}%
  \email{matthias.wilhelm@nbi.ku.dk}

\affiliation{%
Niels Bohr Institute, Copenhagen University,\\
Blegdamsvej 17, 2100 Copenhagen \O{}, Denmark
}%

\begin{abstract}
We initiate the calculation of loop corrections to correlation functions in 4D defect CFTs. 
More precisely,
we consider ${\cal N}=4$ SYM theory with a codimension-one defect separating 
two regions of space, $x_3>0$ and $x_3<0$, 
where the gauge group is $SU(N)$ and $SU(N-k)$, respectively. This set-up is made possible by some of the real scalar fields acquiring a non-vanishing and $x_3$-dependent vacuum expectation value  for $x_3>0$.
The holographic dual is the D3-D5 probe brane system where the D5 brane geometry is $AdS_4\times S^2$
and a background gauge field has $k$ units of flux through the $S^2$.  We diagonalise the mass matrix
of the defect CFT making use of fuzzy-sphere coordinates and we handle the $x_3$-dependence of  the mass terms in 
the 4D Minkowski space propagators by reformulating these as standard massive $AdS_4$ propagators. Furthermore,
we show that only two Feynman diagrams contribute to the one-loop correction to the one-point function of any single-trace operator and we explicitly calculate this correction in the planar limit for the simplest chiral primary. The result of this
calculation is compared to an earlier string-theory computation in a certain double-scaling limit, finding perfect agreement. Finally, we discuss how to generalise our calculation to any single-trace operator, to finite $N$ and to other types of observables such as Wilson loops.
\end{abstract}

\maketitle

\section{Introduction}

Introducing boundaries or defects in conformal field theories leads to novel features concerning
correlation functions~\cite{Cardy:1984bb}.
For instance, one-point functions can be non-vanishing and operators which have different conformal dimensions can have a non-vanishing overlap. Furthermore, such set-ups typically involve additional fields which are confined to the defect and these fields can have   overlaps with the bulk fields.
Via the Karch-Randall idea~\cite{Karch:2000gx}, several examples of defect conformal field theories (dCFTs) with holographic duals have been identified. 

Our focus is on a particular such 4D defect conformal theory, namely
${\cal N}=4$ supersymmetric Yang-Mills ($\mathcal{N}=4$ SYM) theory  with a codimension-one defect separating two regions of space-time
where the gauge group is $SU(N)$ and $SU(N-k)$, respectively~\cite{Nahm:1979yw,Diaconescu:1996rk,Giveon:1998sr,Constable:1999ac}. 
The holographic dual  is the probe D3-D5 brane system
 involving a single
probe D5 brane with geometry $AdS_4\times S^2$ where
a background gauge field has $k$ units of flux on the $S^2$~\cite{Constable:1999ac}.

  A number of one- and two-point functions involving both bulk and defect fields have been  analysed in the zero flux case~\cite{deWolfe:2001pq,Susaki:2004tg,McLoughlin:2005gj,Susaki:2005qn,Okamura:2005cj}, but the 
study of correlation functions in the presence of  flux was only initiated recently. In \cite{Nagasaki:2012re,Kristjansen:2012tn}, tree-level one-point functions of chiral primary operators were calculated. 
For non-protected operators,  
tree-level one-point functions 
are only meaningful for operators which are one-loop eigenstates of the dilatation generator. As is well known, such operators
can be described as  Bethe eigenstates of a certain integrable spin chain~\cite{Minahan:2002ve,Beisert:2003yb}. A systematic method for the calculation of tree-level
one-point functions of non-protected operators was presented in~\cite{deLeeuw:2015hxa,Buhl-Mortensen:2015gfd}, in which the one-point function
was expressed as the overlap between a Bethe eigenstate and a certain matrix product state.  Using the tools of integrable
spin chains, it was possible to derive a closed expression for the one-point function of any operator in the $SU(2)$ sector
valid for any value of the flux, $k$.  The method can be extended to the $SU(3)$ sector, which is a closed sector
at the one-loop level~\cite{deLeeuw:2016umh}.

In the present letter, we initiate the calculation of quantum
corrections to the observables of the above dCFT.
We focus on the one-loop corrections to one-point functions,
but our work also paves the way for the analysis of other types of correlators, of Wilson loops and of computations to higher loop orders.  
The major obstacle in moving on to one-loop level is that the vacuum expectation values (vevs) of the scalar fields, that realise the difference in the gauge group on the two sides of the defect, introduce a highly involved mass matrix, which needs to be diagonalised.
We perform this diagonalisation by making use of fuzzy-sphere coordinates.
Another complication is that the masses in the spectrum all depend on the distance from the defect, which invalidates many
of the traditional field-theoretical methods. We deal with this problem by working with propagators in 
an auxiliary $AdS_4$ space instead of usual 4D Minkowski space propagators. For the one-loop corrections to the one-point functions
of single-trace operators, we find that only two Feynman diagrams contribute and we regulate these using dimensional reduction.
As expected, the dependence of the regulator, $\peps$, drops out and we end up with a finite result. We relegate many details of
our analysis to a forthcoming article~\cite{longpaper}.

\section{The Defect Theory}

Our starting point is the dCFT formulated in~\cite{deWolfe:2001pq}.  It consists of ${\cal N}=4$ SYM theory 
coupled to a 3D hypermultiplet of fundamental fields living on  a codimension-one defect, a set-up
which preserves half of the supersymmetries of ${\cal N}=4$ SYM theory as well as the defect-preserving conformal 
symmetries~\cite{deWolfe:2001pq, Erdmenger:2002ex}.

The action of the system is the sum of the usual ${\cal N}=4$ SYM action and an action describing the self-interactions of the
defect fields and their couplings to the fields of ${\cal N}=4$ SYM theory.
It will turn out that the defect fields play no role at the loop order we consider.
We use the ${\cal N}=4$ SYM action in the  form
\begin{multline}
 \label{eq: SYM-action}
 S_{{\cal N}=4}=\frac{2}{g_\YM^2}\int \de^4x\tr\biggl[ -\frac{1}{4}F_{\mu\nu}F^{\mu\nu}-\frac{1}{2}\cder_\mu\scal_i\cder^\mu\scal_i\\+\frac{i}{2}\aferm\Gamma^\mu\cder_\mu\ferm +\frac{1}{2}\aferm\Gamma^i\comm{\scal_i}{\ferm}+\frac{1}{4}\comm{\scal_i}{\scal_j}\comm{\scal_i}{\scal_j}\biggr]\eqncom
\end{multline}
where $ F_{\mu\nu}= \partial_\mu A_\nu-\partial_\nu A_\mu-i\comm{A_\mu}{A_\nu}$, 
  $\cder_\mu= \partial_\mu-i\comm{A_\mu}{\cdot}$ and  $\{\Gamma_\mu, \Gamma_i\}$ are the 10-dimensional gamma matrices in the Majorana-Weyl representation.
A situation where the defect separates two regions of space with different ranks of the gauge group is realised by the so-called fuzzy-funnel solution \cite{Constable:1999ac}, in which three of
the scalar fields of ${\cal N}=4$ SYM theory acquire a non-vanishing vev on one side of the defect. If the codimension-one defect is placed
at $x_3=0$, the vevs of the scalar fields take the form
\begin{equation}
\label{eq: classical solution}
\langle\phi_i\rangle_{\text{tree}}= \scalc_i=-\frac{1}{x_3} t_i\oplus 0_{(N-k)\times(N-k)}\eqncom \hspace{0.5cm} x_3>0 \eqncom
\end{equation}
where $i=1,2,3$ and where all other classical fields are set to zero.
Here, $t_1,t_2$ and $t_3$ are generators of the $SU(2)$ Lie algebra in the $k$-dimensional irreducible representation. 
With this set-up,
the gauge group is (broken) $SU(N)$ for $x_3>0$ and $SU(N-k)$ for $x_3<0$. 

To perform perturbative calculations, we expand the
scalar fields around their classical values 
\begin{equation}
\phi_i=\scalc_i+\scalq_i\eqncom \hspace{0.5cm} i=1,2,3\eqndot
\end{equation}
Furthermore, we fix the gauge by adding the following term to the action \eqref{eq: SYM-action}:
\begin{equation}
S_{\text{gf}}= -\frac{1}{2}\frac{2}{g_\YM^2}\int \de^4x\tr(G^2)\eqncom \quad
 G=\partial_\mu A^\mu+i\comm{\scalq_i}{\scalc_i}\eqndot
\end{equation}
This also cancels an unwanted term linear in the derivative, which arises when expanding \eqref{eq: SYM-action} around the classical solution.

The resulting gauge-fixed action is 
\begin{equation}
S_{{\cal N}=4}+S_{\text{gf}} + S_{\text{ghost}} =
S_{\text{kin}} + S_{\text{m}} + S_{\text{cubic}} + S_{\text{quartic}}\eqncom
\end{equation}
where the Gaussian part consists of the kinetic terms
\begin{multline}
  S_{\text{kin}}
    = \frac{2}{g_\YM^2}\int \de^4 x \tr\biggl[
        \frac{1}{2}A_\mu \partial_\nu\partial^\nu A^\mu
        +\frac{1}{2}\scalq_i \partial_\nu\partial^\nu \scalq_i\\
        +\frac{i}{2}\aferm\Gamma^\mu\partial_\mu\ferm
        +\bar{c}\,\partial_\mu\partial^\mu c\biggr] \eqncom
\end{multline}
and the mass terms
\begin{multline}
  \label{eq: mass}
 \!\!\!\! S_{\text{m}}
    = \frac{2}{g_\YM^2}\int \de^4 x \tr\biggl[
        \frac{1}{2}\comm{\scalc_i}{\scalc_j}\comm{\scalq_i}{\scalq_j}
        +\frac{1}{2}\comm{\scalc_i}{\scalq_j}\comm{\scalc_i}{\scalq_j}\\
        +\frac{1}{2}\comm{\scalc_i}{\scalq_j}\comm{\scalq_i}{\scalc_j}
        +\frac{1}{2}[\scalc_i,\scalq_i][\scalc_j,\scalq_j]
        +\frac{1}{2}[A_\mu,\scalc_i][A^\mu,\scalc_i]\\
        +2i[A^\mu,\scalq_i]\partial_\mu\scalc_i
        +\frac{1}{2}\aferm\Gamma^i[\scalc_i,\ferm]
        -\bar{c}\,[\scalc_i,[\scalc_i,c]]\biggr]\eqndot
\end{multline}
The interactions are given by the cubic vertices
\begin{multline}
        \label{eq: cubic}
  S_{\text{cubic}}
    = \frac{2}{g_\YM^2}\int \de^4 x \tr\biggl[
        i[A^\mu,A^\nu]\partial_\mu A_\nu 
        +i[A^\mu,\scalq_i]\partial_\mu\scalq_i\\
        +[\scalc_i,\scalq_j][\scalq_i,\scalq_j]
        +[A_\mu,\scalc_i][A^\mu,\scalq_i]
        +\frac{1}{2}\aferm\Gamma^\mu[A_\mu,\ferm]\\
        +\frac{1}{2}\aferm\Gamma^i[\scalq_i,\ferm]
        +i(\partial_\mu\bar{c})[A^\mu,c]
        -\bar{c}\,[\scalc_i,[\scalq_i,c]]\biggr]\eqncom
\end{multline}
plus a number of standard quartic vertices which will not play any role. 
Here, $c$ and its conjugate $\bar{c}$ are fermionic (but Lorentz scalar) ghost fields.

Note that \eqref{eq: mass} are not usual mass terms, as they depend on the classical solution $\scalc_i$ and hence on the distance $x_3$ to the defect.   Moreover, they are non-diagonal in both flavour and colour.  Not all flavours mix, though. The
colour components of the gauge field $A_0$ only mix among themselves and not with colour components of any other fields. The same is true for the colour components of $A_1$ and $A_2$ as well as for the colour components of the scalars $\scalq_4,\scalq_5$ and $\scalq_6$ and the ghosts. For the remaining bosonic fields $\scalq_1$,
$\scalq_2$, $\scalq_3$, $A_3$ and the original fermions, the mixing problem is more complicated and involves both flavour and colour. We
find that the mixing problem can be completely solved by making use of fuzzy-sphere coordinates. 
We present the eigenvalues and corresponding multiplicities in table~\ref{tab:spectrum}, while deferring the
detailed derivation to a forthcoming
paper~\cite{longpaper}. Notice that we have left out the factor $1/x_3$ in table~\ref{tab:spectrum}, which multiplies all masses 
in the diagonalised  action.
For the bosonic fields, the mass eigenvalues are 
 expressed in terms of 
\begin{equation}
\label{eq: def nu}
 \nu=\sqrt{m^2+\frac{1}{4}} \eqndot
\end{equation}
The mass matrix of the fermions $\psi$ has positive as well as negative eigenvalues. In order to obtain the canonical form of the action with positive masses, the sign of the latter can be changed via a chiral rotation of the fermions.

\begin{table}[t]
\begin{tabular}{c|c|c|c}
Multiplicity & $\nu(\scalq_{4,5,6},A_{0,1,2},c)$ & $m(\psi_{1,2,3,4})$ & $\nu(\scalq_{1,2,3},A_3)$ \\ \hline
$\ell +1$ & $\ell+\frac{1}{2}$ & $-\ell$ & $\ell-\frac{1}{2}$ \\
$\ell$ & $\ell+\frac{1}{2}$ & $\ell+1$ & $\ell+\frac{3}{2}$ \\
$(k+1)(N-k)$ & $\frac{k}{2}$ & $-\frac{k-1}{2}$ & $\frac{k-2}{2}$ \\
$(k-1)(N-k)$ & $\frac{k}{2}$ & $\frac{k+1}{2}$ & $\frac{k+2}{2}$ \\
$(N-k)(N-k)$ & $\frac{1}{2}$ & $0$ & $\frac{1}{2}$
\end{tabular}
\caption{\label{tab:spectrum}Masses and multiplicities of the different fields with $\ell=1,\ldots,k-1$, partially given in terms of $\nu$ defined in \eqref{eq: def nu}.} 
\end{table}

Once we have diagonalised the mass matrix, the propagators are obtained in the usual way. 
Hence, a scalar propagator $K(x,y)$ is the solution to
\begin{equation}
  \left(-\partial_\mu\partial^\mu+\frac{m^2}{(x_3)^2}\right)K(x,y) = \frac{g_\YM^2}{2}\delta(x-y)
  \eqncom
  \label{eq:prop-diff-eq}
  \end{equation}
where the derivatives are with respect to $x$.  
  If one compares this to the definition of the propagator $K_{\text{AdS}}(x,y)$ of a scalar in $AdS_4$ with mass $\tilde{m}$ 
 \begin{equation}
  (-\nabla_\mu\nabla^\mu + \tilde{m}^2)K_{\text{AdS}}(x,y)
    = \frac{\delta(x-y)}{\sqrt g}\eqncom
\end{equation}
with the metric of $AdS_4$ given as $g_{\mu\nu} = (x_3)^{-2}\,\eta_{\mu\nu}$,
 one concludes that
\begin{equation}
  K(x,y) 
  = \frac{g_\YM^2}{2}\frac{K_{\text{AdS}}(x,y)}{x_3y_3}\eqncom
\end{equation}
with the identification $\tilde{m}^2=m^2-2.$
We notice the satisfying fact that none of the scalar masses in table~\ref{tab:spectrum} leads to a violation of the
Breitenlohner-Freedman (BF) bound~\cite{Breitenlohner:1982jf}, since $\tilde{m}^2\geq -\frac{9}{4}$,
which is exactly the BF bound for $AdS_4$. The bound is only saturated in the special case
$k=2$.
Closed expressions  for
$K_{\text{AdS}}(x,y)$ in terms of hypergeometric functions can be found in
the literature, see e.g.~\cite{Allen:1985wd,Camporesi91}. A representation which is particularly useful for our purpose is~\cite{Liu:1998ty}
\begin{equation}
\begin{aligned}
K(x,y)
={}&\frac{g_\YM^2\sqrt{x_3 y_3}}{2} \\ &\int \frac{\de^3 \vec{k}}{(2\pi)^3}\e^{i \vec{k}\cdot(\vec{x}-\vec{y})} I_{\nu}(|\vec{k}| x_3^<) K_{\nu}(|\vec{k}| x_3^>)\eqncom
\end{aligned}
\end{equation}
where $I_\nu$ and $K_\nu$ are modified Bessel functions with $\nu$ given in \eqref{eq: def nu} and
with $x_3^<$ ($x_3^>$) the smaller (larger) of $x_3$ and $y_3$. Furthermore, $\vec{k}=(k_0,k_1,k_2)$ denotes the directions parallel to the defect. For the propagators of the spinor fields, one finds by
similar considerations
\begin{equation}
  K^F(x,y) 
   = \frac{g_\YM^2}{2}\frac{K^F_{\text{AdS}}(x,y)}{(x_3)^{3/2}(y_3)^{3/2}}\eqncom
\end{equation}
this time with $\tilde{m}_F={m}_F$.
For more details, we refer to~\cite{longpaper}. Our considerations are an elaboration of the statement already made 
 in~\cite{Nagasaki:2011ue} that the mass terms could be rendered position independent by performing a Weyl transformation to  $AdS_4$ space.

\section{One-point functions}
With the classical fields given by~\eqref{eq: classical solution}, single-trace operators built from the scalar fields $\phi_1, \phi_2$
and $\phi_3$ will have non-vanishing one-point functions on one side of the defect, $x_3>0$, already at tree level with the expected space-time dependence~\cite{Cardy:1984bb}:
\begin{equation}
\langle {\cal O}_{\Delta}\rangle=\frac{C}{x_3^{\Delta}}\eqncom
\end{equation}
where $C$ is a constant and $\Delta$ denotes the scaling dimension of ${\cal O}$.
\begin{figure}[t]
\centering
 \subfigure[]{
\centering
  \includegraphics{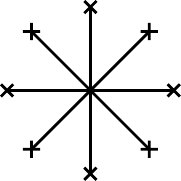}%
\begin{picture}(0, 0)
\put(-0.5,0.5){\circle*{0.1}}
\end{picture}
  \label{subfig: tree}
} \qquad 
 \subfigure[]{
\centering
  \includegraphics{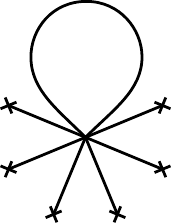}%
\begin{picture}(0, 0)
\put(-0.4725,0.47){\circle*{0.1}}
\end{picture}
  \label{subfig: tadpole}
} \qquad 
 \subfigure[]{
\centering
  \includegraphics{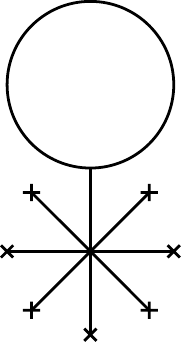}%
\begin{picture}(0, 0)
\put(-0.5,0.5){\circle*{0.1}}
\end{picture}
  \label{subfig: lolipop}
}
\caption{Tree-level \subref{subfig: tree} and one-loop (\subref{subfig: tadpole} tadpole and \subref{subfig: lolipop} lollipop) contributions to one-point functions. A cross stands for the insertion of the classical solution, while the operator is depicted as a dot.}
\label{fig: one-point functions}
\end{figure}
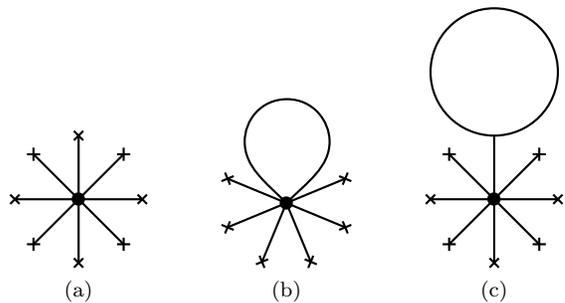
For simplicity,
we illustrate our method by
considering operators which do not get corrected (in the theory without the defect), i.e.\ the chiral primaries of ${\cal N}=4$ SYM theory. 
Furthermore,  we will consider the simplest such operator
\begin{equation}
\label{eq: operator}
  \calO(x) = \tr(Z^L)(x)\eqncom\quad Z(x)=\scal_3(x)+i\phi_{6}(x) \eqndot
\end{equation}
At tree level, the one-point function of $\calO$ is given by inserting \eqref{eq: classical solution} into \eqref{eq: operator}, as depicted in figure \ref{subfig: tree}. This yields \cite{deLeeuw:2015hxa}
\begin{equation}
\begin{aligned}
 \langle\calO\rangle_{\text{tree-level}} =
-\frac{2}{x_3^{L}(L+1)}B_{L+1}\left(\frac{1-k}{2}\right)
\end{aligned}
\end{equation}
for $L$ even and vanishes when $L$ is odd. Here, $B_{L+1}(u)$ are the Bernoulli polynomials.
We have not divided by the norm of the two-point function, since this normalisation factor will not play any role in our analysis \footnote{It can be found e.g.\ in~\cite{deLeeuw:2015hxa}}.

At one-loop order, there are two possible Feynman diagrams which
we depict  in figures \ref{subfig: tadpole} and \ref{subfig: lolipop} and denote  as the tadpole and the lollipop diagram.
Symbolically, the tadpole contribution looks like
\begin{equation}
  \langle\calO\rangle_{\text{1-loop},\text{tad}} \sim \frac{1}{x_3^{L-2}}\sum_m K(x,x)\eqndot
\end{equation}
The sum is over the spectrum of the relevant (scalar) modes, and we have omitted
the similarity transformations that change between the original and
mass-diagonal basis.
Symbolically, the lollipop diagram contributes as follows:
\begin{equation}
  \langle\calO\rangle_{\text{1-loop},\text{lol}} \sim \frac{g_\YM^{-2}}{x_3^{L-1}}\sum_{m_1,m_2} 
    \int \de^4 y\, K_1(x,y)V K_2(y,y)\eqndot
\end{equation}
Here, $m_1$ ranges only over bosonic modes, whereas $m_2$ also includes fermions.
The vertex factor $V$ is $\propto 1/y_3$ for scalars, gluons and ghosts in the loop
but just a number for fermions.
Again, we have neglected many factors.  One can 
convince oneself that the quartic interaction terms do not contribute at one-loop order.
Likewise,  the defect fields do not play any role at one-loop order; the only way a defect field could contribute at one-loop order would involve a tadpole diagram of the 3D theory living on the defect, which vanishes due to conformal invariance.

Both the scalar and the fermion loop are divergent and require regularisation. We regulate
using dimensional reduction~\cite{Siegel:1979wq} in the $d=3-2\peps$ dimensions parallel to the defect
and show that all dependence on the regulator, $\peps$, cancels out in the final result. This constitutes a strong consistency
check of our calculations.
For the scalar loop $K(x,x)$ with $m \neq 0$,
dimensional regularisation leads to \cite{longpaper}
\begin{equation}
  \label{eq:scalar-loop}
\begin{aligned}  
  K(x,x)
    = 
    \frac{g_\YM^2}{2}
 \frac{1}{16\pi^2\, x_3^2} \biggl( m^2 \biggl[- \frac{1}{\peps} -\log(4\pi)+\gammaE \qquad \\-2\log (x_3)+ 2\digamma(\nu+\tfrac{1}{2})-1  \biggr]-1\biggr)
    \eqndot
\end{aligned}    
\end{equation}
Here, $\gammaE$ is the Euler-Mascheroni constant and $\Psi$ is the Euler digamma function.
The fermion loop 
in dimensional regularisation reads 
%
\begin{equation}
  \label{eq:fermion-loop}
\begin{aligned}
   &\tr K^F(x,x) =\\&\frac{m}{|m|} \frac{g_\YM^2}{2}\,\frac{1}{4\pi^2x_3^3}\Bigg[ |m|^3+|m|^2-3|m|-1   
+|m|(|m|^2-1)\\&\times\left(-\frac{1}{\peps} -\log(4\pi)+\gammaE-2 \log (x_3) +2\digamma(|m|)-2\right)\Bigg] \eqncom
      \end{aligned}
\end{equation}
where the sign of the mass, $m/|m|$, stems from the aforementioned chiral rotation of the fermions.

In the present letter, we shall restrict ourselves to calculating the large-$N$ contribution to the one-point function. The evaluation
of the finite-$N$ contribution poses no conceptual problems but involves colour components of the fields which can be
ignored in the large-$N$ limit.  We refer to~\cite{longpaper}
for a more detailed discussion.
In the large-$N$ limit, only tadpole
diagrams where the tadpole connects neighbouring fields contribute and there are $L$ such terms. The excitations which run
in the loop can either be $\scalq_3$ or $\scalq_6$ and both of the associated contributions can be calculated explicitly.  This leads
to the following result, valid for even $L$
\begin{equation}\label{tadresult}
\langle {\cal O}\rangle_{\text{1-loop},\text{tad}}= -\frac{\lambda} {16 \pi^2}\,\frac{2L}{x_3^{L}(L-1)}\, 
  B_{L-1}\left(\frac{1-k}{2}\right) \eqndot
\end{equation} 
The contribution vanishes for odd $L$.

The evaluation of the contribution from the lollipop diagram is considerably more involved. First, the large-$N$ limit only constrains the type of colour components for the fields which run in the loop and not for the fields which run in the stick. Second, one needs to repeatedly use the similarity transformation which relates the mass eigenstates to the various field components.
Finally, 
the use of a supersymmetry-preserving renormalisation scheme is crucial.
Assembling the numerous contributions, we find
that the lollipop contribution vanishes: 
\begin{equation}\label{lolresultfirst}
 \langle {\cal O}\rangle_{\text{1-loop},\text{lol}} =
 0
\eqndot
\end{equation}
For details on the calculation, in particular on the similarity transformation to the mass eigenbasis which features heavily in it, see \cite{longpaper}. Notice that in both~(\ref{tadresult}) and~(\ref{lolresultfirst}) all dependence on the regulator $\peps$ has cancelled out and so have the various logarithms and the Euler-Mascheroni constant.
Note also that the contribution of the lollipop diagram can be equivalently obtained from 
the one-loop correction to the vev of the scalars, $\langle\phi_i\rangle_{\text{1-loop}}$, 
which equally vanishes.

\section{Comparison to string theory \label{comparison}}

The present calculations open a new possibility of comparing results between gauge and string theory
with less  (super) symmetries. In particular, we have at our disposal a novel parameter $k$.
In \cite{Nagasaki:2011ue,Nagasaki:2012re}, it was suggested to consider a 
limit which consists
in letting $N\rightarrow \infty$ and subsequently  $k\rightarrow \infty$ (but $k\ll N$) while keeping $\lambda/k^2\ll1$. 

In the string-theory
language, the $N\rightarrow \infty$ limit  eliminates string interactions and the limit $\lambda\rightarrow \infty$ 
justifies a supergravity treatment.  The string configuration dual to a one-point
function is that of
a string stretching from the boundary of $AdS_5$ (more precisely from the insertion point of the dual  gauge-theory
operator) and ending on the $D5$-brane in the interior of $AdS_5\times S^5$.  In the case of a chiral primary, the string
can be considered point-like and the one-point function can be computed using a variant of the Witten prescription~\cite{Nagasaki:2012re, Witten:1998qj,Buhl-Mortensen:2015gfd}. In the limit described above, the result organises into
a power series expansion in $\lambda/k^2$. This led the authors of \cite{Nagasaki:2012re} to suggest that the result
might match the result of a perturbative gauge-theory computation, which, however, would require
that the gauge-theory perturbative result would likewise organise itself into a power series expansion in $\lambda/k^2$. This
idea is very reminiscent of the BMN idea~\cite{Berenstein:2002jq} fostered in connection with the study of the 
spectral problem of ${\cal N}=4$ SYM theory. Here, another quantum number, $J$, which had the
interpretation of an $S^5$ angular momentum of a spinning string, was considered large as well as  $\lambda$
while  $\lambda/J^2$ was assumed to be finite. In the BMN case, it eventually turned out that starting
at four-loop order the perturbative gauge-theory expansion of anomalous dimensions did not organise itself into 
powers of $\lambda/J^2$~\cite{Beisert:2006ez,Bern:2006ew,Cachazo:2006az}.

The authors of~\cite{Nagasaki:2012re} showed that the leading term in
the $\frac{\lambda}{k^2}$ expansion matches the tree-level gauge-theory result. Their supergravity result, however, also implies a prediction for the one-loop gauge-theory correction to the one-point function. The chiral primary of length $L$ considered in~\cite{Nagasaki:2012re} is
not the same as~\eqref{eq: operator} but has a non-vanishing projection onto the latter. 
Thus, the ratio of the next-to-leading-order term and the leading-order term in $\lambda/k^2$ should match the ratio between our one-loop and 
tree-level result.  The prediction for this ratio following from~\cite{Nagasaki:2012re} reads
\begin{equation}
\left.\frac{\langle {\cal O}\rangle_{\text{1-loop}}}{\langle {\cal O}\rangle_{\text{tree-level}}}\right|_{\text{string}}=
\frac{\lambda}{4\pi^2 k^2} \frac{L(L+1)}{L-1}\eqndot
\end{equation}
For the tadpole diagram and the vanishing lollipop diagram, we find
\begin{equation}
\left.\frac{\langle {\cal O}\rangle_{\text{1-loop}}}{\langle {\cal O}\rangle_{\text{tree-level}}}\right|_{\text{gauge}}=
\frac{\lambda}{4\pi^2 k^2}\left( \frac{L(L+1)}{L-1} + O(k^{-2})\right)\eqncom
\end{equation}
which is identical to the supergravity result in the double-scaling limit.
This match provides a highly non-trivial check of the gauge-gravity duality in the case of partially broken supersymmetry as well as conformal symmetry!

\section{Conclusions \& Outlook}

With the present work, we have laid the foundation for a detailed analysis of a class of dCFTs based on ${\cal N}=4$ SYM theory, which have holographic duals involving background gauge fields with flux. 
The flux, which is related to the difference in rank of the gauge group on the two sides of a defect, constitutes an interesting extra tunable parameter of the 
AdS/dCFT set-up. Its presence severely complicates the field-theoretical analysis since some of the scalar fields
of ${\cal N}=4$ SYM theory acquire non-vanishing and space-time-dependent vevs, which leads to a highly
non-trivial mixing  both at the flavour and at the colour level. We have
solved this mixing problem and diagonalised the mass matrix of the theory. In addition, we have shown how to trade Minkowski
space propagators with space-time-dependent mass terms for $AdS$ space propagators with standard mass terms. 
With these two steps accomplished, the perturbative calculation of observables in the dCFT can be carried out by
standard methods. We illustrated this by calculating the planar one-loop correction to the one-point function of the chiral primary
operator $\tr(Z^L)$.  
In a certain double-scaling limit, our gauge-theory result perfectly agrees with an earlier prediction for the same quantity from string theory. 
This provides a strong test of the AdS/dCFT duality at quantum level.

Our analysis can be extended in numerous directions.
First, it is straightforward to extend the calculation to finite $N$. Second, the calculation can be
generalised to any  operator built  of scalars.  This might reveal interesting novel structures,
as integrability has recently shown its face in the calculation of  tree-level one-point functions 
in the $SU(2)$ sector~\cite{deLeeuw:2015hxa,Buhl-Mortensen:2015gfd}.  
It would also be interesting to investigate the types of correlators special to dCFTs such as two-point functions
between bulk operators with different conformal dimensions and two-point functions involving both bulk and defect fields. Moreover,
one could envision going to higher loop orders where
presumably starting from two-loop order the defect fields would come into play and present further challenges.
Finally, some simple examples of Wilson loops in the present defect set-up were considered in~\cite{Nagasaki:2011ue}, where a tree-level computation was carried out on
the field-theory side and compared to a supergravity computation. As for one-point functions, 
agreement was observed between the tree-level and the supergravity result in the double-scaling limit described above. 
It would be interesting to address this at one-loop order.


\begin{acknowledgments}
\paragraph{Acknowledgements.}
We thank S.\ Caron-Huot, G.\ Semenoff and K.\ Zarembo for very useful discussions.
I.B.-M.,  M.d.L., C.K.\  and M.W.\ were  supported  in  part  by  FNU  through
grants number DFF-1323-00082 and DFF-4002-00037. A.C.I.\ acknowledges the support of the ERC Advanced Grant 291092.
All authors acknowledge the kind hospitality of NORDITA during the program ``Holography and Dualities,'' where parts of this work were carried out.
\end{acknowledgments}

\bibliography{lettershortened}

\begin{thebibliography}{32}%
\makeatletter
\providecommand \@ifxundefined [1]{%
 \@ifx{#1\undefined}
}%
\providecommand \@ifnum [1]{%
 \ifnum #1\expandafter \@firstoftwo
 \else \expandafter \@secondoftwo
 \fi
}%
\providecommand \@ifx [1]{%
 \ifx #1\expandafter \@firstoftwo
 \else \expandafter \@secondoftwo
 \fi
}%
\providecommand \natexlab [1]{#1}%
\providecommand \enquote  [1]{``#1''}%
\providecommand \bibnamefont  [1]{#1}%
\providecommand \bibfnamefont [1]{#1}%
\providecommand \citenamefont [1]{#1}%
\providecommand \href@noop [0]{\@secondoftwo}%
\providecommand \href [0]{\begingroup \@sanitize@url \@href}%
\providecommand \@href[1]{\@@startlink{#1}\@@href}%
\providecommand \@@href[1]{\endgroup#1\@@endlink}%
\providecommand \@sanitize@url [0]{\catcode `\\12\catcode `\$12\catcode
  `\&12\catcode `\#12\catcode `\^12\catcode `\_12\catcode `\%12\relax}%
\providecommand \@@startlink[1]{}%
\providecommand \@@endlink[0]{}%
\providecommand \url  [0]{\begingroup\@sanitize@url \@url }%
\providecommand \@url [1]{\endgroup\@href {#1}{\urlprefix }}%
\providecommand \urlprefix  [0]{URL }%
\providecommand \Eprint [0]{\href }%
\providecommand \doibase [0]{http://dx.doi.org/}%
\providecommand \selectlanguage [0]{\@gobble}%
\providecommand \bibinfo  [0]{\@secondoftwo}%
\providecommand \bibfield  [0]{\@secondoftwo}%
\providecommand \translation [1]{[#1]}%
\providecommand \BibitemOpen [0]{}%
\providecommand \bibitemStop [0]{}%
\providecommand \bibitemNoStop [0]{.\EOS\space}%
\providecommand \EOS [0]{\spacefactor3000\relax}%
\providecommand \BibitemShut  [1]{\csname bibitem#1\endcsname}%
\let\auto@bib@innerbib\@empty
\bibitem [{\citenamefont {Cardy}(1984)}]{Cardy:1984bb}%
  \BibitemOpen
  \bibfield  {author} {\bibinfo {author} {\bibfnamefont {John~L.}\ \bibnamefont
  {Cardy}},\ }\bibfield  {title} {\enquote {\bibinfo {title} {{Conformal
  Invariance and Surface Critical Behavior}},}\ }\href {\doibase
  10.1016/0550-3213(84)90241-4} {\bibfield  {journal} {\bibinfo  {journal}
  {Nucl. Phys.}\ }\textbf {\bibinfo {volume} {B240}},\ \bibinfo {pages}
  {514--532} (\bibinfo {year} {1984})}\BibitemShut {NoStop}%
\bibitem [{\citenamefont {Karch}\ and\ \citenamefont
  {Randall}(2001)}]{Karch:2000gx}%
  \BibitemOpen
  \bibfield  {author} {\bibinfo {author} {\bibfnamefont {Andreas}\ \bibnamefont
  {Karch}}\ and\ \bibinfo {author} {\bibfnamefont {Lisa}\ \bibnamefont
  {Randall}},\ }\bibfield  {title} {\enquote {\bibinfo {title} {{Open and
  closed string interpretation of SUSY CFT's on branes with boundaries}},}\
  }\href {\doibase 10.1088/1126-6708/2001/06/063} {\bibfield  {journal}
  {\bibinfo  {journal} {JHEP}\ }\textbf {\bibinfo {volume} {06}},\ \bibinfo
  {pages} {063} (\bibinfo {year} {2001})},\ \Eprint
  {http://arxiv.org/abs/hep-th/0105132} {arXiv:hep-th/0105132 [hep-th]}
  \BibitemShut {NoStop}%
\bibitem [{\citenamefont {Nahm}(1980)}]{Nahm:1979yw}%
  \BibitemOpen
  \bibfield  {author} {\bibinfo {author} {\bibfnamefont {W.}~\bibnamefont
  {Nahm}},\ }\bibfield  {title} {\enquote {\bibinfo {title} {{A Simple
  Formalism for the BPS Monopole}},}\ }\href {\doibase
  10.1016/0370-2693(80)90961-2} {\bibfield  {journal} {\bibinfo  {journal}
  {Phys. Lett.}\ }\textbf {\bibinfo {volume} {B90}},\ \bibinfo {pages}
  {413--414} (\bibinfo {year} {1980})}\BibitemShut {NoStop}%
\bibitem [{\citenamefont {Diaconescu}(1997)}]{Diaconescu:1996rk}%
  \BibitemOpen
  \bibfield  {author} {\bibinfo {author} {\bibfnamefont {Duiliu-Emanuel}\
  \bibnamefont {Diaconescu}},\ }\bibfield  {title} {\enquote {\bibinfo {title}
  {{D-branes, monopoles and Nahm equations}},}\ }\href {\doibase
  10.1016/S0550-3213(97)00438-0} {\bibfield  {journal} {\bibinfo  {journal}
  {Nucl. Phys.}\ }\textbf {\bibinfo {volume} {B503}},\ \bibinfo {pages}
  {220--238} (\bibinfo {year} {1997})},\ \Eprint
  {http://arxiv.org/abs/hep-th/9608163} {arXiv:hep-th/9608163 [hep-th]}
  \BibitemShut {NoStop}%
\bibitem [{\citenamefont {Giveon}\ and\ \citenamefont
  {Kutasov}(1999)}]{Giveon:1998sr}%
  \BibitemOpen
  \bibfield  {author} {\bibinfo {author} {\bibfnamefont {Amit}\ \bibnamefont
  {Giveon}}\ and\ \bibinfo {author} {\bibfnamefont {David}\ \bibnamefont
  {Kutasov}},\ }\bibfield  {title} {\enquote {\bibinfo {title} {{Brane dynamics
  and gauge theory}},}\ }\bibfield  {booktitle} {\emph {\bibinfo {booktitle}
  {{Strings. Proceedings, International Conference, Strings'97, Amsterdam,
  Netherlands, June 16-21, 1997}}},\ }\href {\doibase
  10.1103/RevModPhys.71.983} {\bibfield  {journal} {\bibinfo  {journal} {Rev.
  Mod. Phys.}\ }\textbf {\bibinfo {volume} {71}},\ \bibinfo {pages} {983--1084}
  (\bibinfo {year} {1999})},\ \Eprint {http://arxiv.org/abs/hep-th/9802067}
  {arXiv:hep-th/9802067 [hep-th]} \BibitemShut {NoStop}%
\bibitem [{\citenamefont {Constable}\ \emph {et~al.}(2000)\citenamefont
  {Constable}, \citenamefont {Myers},\ and\ \citenamefont
  {Tafjord}}]{Constable:1999ac}%
  \BibitemOpen
  \bibfield  {author} {\bibinfo {author} {\bibfnamefont {Neil~R.}\ \bibnamefont
  {Constable}}, \bibinfo {author} {\bibfnamefont {Robert~C.}\ \bibnamefont
  {Myers}}, \ and\ \bibinfo {author} {\bibfnamefont {Oyvind}\ \bibnamefont
  {Tafjord}},\ }\bibfield  {title} {\enquote {\bibinfo {title} {{The
  Noncommutative bion core}},}\ }\href {\doibase 10.1103/PhysRevD.61.106009}
  {\bibfield  {journal} {\bibinfo  {journal} {Phys. Rev.}\ }\textbf {\bibinfo
  {volume} {D61}},\ \bibinfo {pages} {106009} (\bibinfo {year} {2000})},\
  \Eprint {http://arxiv.org/abs/hep-th/9911136} {arXiv:hep-th/9911136 [hep-th]}
  \BibitemShut {NoStop}%
\bibitem [{\citenamefont {DeWolfe}\ \emph {et~al.}(2002)\citenamefont
  {DeWolfe}, \citenamefont {Freedman},\ and\ \citenamefont
  {Ooguri}}]{deWolfe:2001pq}%
  \BibitemOpen
  \bibfield  {author} {\bibinfo {author} {\bibfnamefont {Oliver}\ \bibnamefont
  {DeWolfe}}, \bibinfo {author} {\bibfnamefont {Daniel~Z.}\ \bibnamefont
  {Freedman}}, \ and\ \bibinfo {author} {\bibfnamefont {Hirosi}\ \bibnamefont
  {Ooguri}},\ }\bibfield  {title} {\enquote {\bibinfo {title} {{Holography and
  defect conformal field theories}},}\ }\href {\doibase
  10.1103/PhysRevD.66.025009} {\bibfield  {journal} {\bibinfo  {journal} {Phys.
  Rev.}\ }\textbf {\bibinfo {volume} {D66}},\ \bibinfo {pages} {025009}
  (\bibinfo {year} {2002})},\ \Eprint {http://arxiv.org/abs/hep-th/0111135}
  {arXiv:hep-th/0111135 [hep-th]} \BibitemShut {NoStop}%
\bibitem [{\citenamefont {Susaki}\ \emph
  {et~al.}(2005{\natexlab{a}})\citenamefont {Susaki}, \citenamefont
  {Takayama},\ and\ \citenamefont {Yoshida}}]{Susaki:2004tg}%
  \BibitemOpen
  \bibfield  {author} {\bibinfo {author} {\bibfnamefont {Yoshiaki}\
  \bibnamefont {Susaki}}, \bibinfo {author} {\bibfnamefont {Yastoshi}\
  \bibnamefont {Takayama}}, \ and\ \bibinfo {author} {\bibfnamefont {Kentaroh}\
  \bibnamefont {Yoshida}},\ }\bibfield  {title} {\enquote {\bibinfo {title}
  {{Open semiclassical strings and long defect operators in AdS / dCFT
  correspondence}},}\ }\href {\doibase 10.1103/PhysRevD.71.126006} {\bibfield
  {journal} {\bibinfo  {journal} {Phys. Rev.}\ }\textbf {\bibinfo {volume}
  {D71}},\ \bibinfo {pages} {126006} (\bibinfo {year} {2005}{\natexlab{a}})},\
  \Eprint {http://arxiv.org/abs/hep-th/0410139} {arXiv:hep-th/0410139 [hep-th]}
  \BibitemShut {NoStop}%
\bibitem [{\citenamefont {McLoughlin}\ and\ \citenamefont
  {Swanson}(2005)}]{McLoughlin:2005gj}%
  \BibitemOpen
  \bibfield  {author} {\bibinfo {author} {\bibfnamefont {Tristan}\ \bibnamefont
  {McLoughlin}}\ and\ \bibinfo {author} {\bibfnamefont {Ian}\ \bibnamefont
  {Swanson}},\ }\bibfield  {title} {\enquote {\bibinfo {title} {{Open string
  integrability and AdS/CFT}},}\ }\href {\doibase
  10.1016/j.nuclphysb.2005.06.014} {\bibfield  {journal} {\bibinfo  {journal}
  {Nucl. Phys.}\ }\textbf {\bibinfo {volume} {B723}},\ \bibinfo {pages}
  {132--162} (\bibinfo {year} {2005})},\ \Eprint
  {http://arxiv.org/abs/hep-th/0504203} {arXiv:hep-th/0504203 [hep-th]}
  \BibitemShut {NoStop}%
\bibitem [{\citenamefont {Susaki}\ \emph
  {et~al.}(2005{\natexlab{b}})\citenamefont {Susaki}, \citenamefont
  {Takayama},\ and\ \citenamefont {Yoshida}}]{Susaki:2005qn}%
  \BibitemOpen
  \bibfield  {author} {\bibinfo {author} {\bibfnamefont {Yoshiaki}\
  \bibnamefont {Susaki}}, \bibinfo {author} {\bibfnamefont {Yastoshi}\
  \bibnamefont {Takayama}}, \ and\ \bibinfo {author} {\bibfnamefont {Kentaroh}\
  \bibnamefont {Yoshida}},\ }\bibfield  {title} {\enquote {\bibinfo {title}
  {{Integrability and higher loops in AdS/dCFT correspondence}},}\ }\href
  {\doibase 10.1016/j.physletb.2005.07.058} {\bibfield  {journal} {\bibinfo
  {journal} {Phys. Lett.}\ }\textbf {\bibinfo {volume} {B624}},\ \bibinfo
  {pages} {115--124} (\bibinfo {year} {2005}{\natexlab{b}})},\ \Eprint
  {http://arxiv.org/abs/hep-th/0504209} {arXiv:hep-th/0504209 [hep-th]}
  \BibitemShut {NoStop}%
\bibitem [{\citenamefont {Okamura}\ \emph {et~al.}(2006)\citenamefont
  {Okamura}, \citenamefont {Takayama},\ and\ \citenamefont
  {Yoshida}}]{Okamura:2005cj}%
  \BibitemOpen
  \bibfield  {author} {\bibinfo {author} {\bibfnamefont {Keisuke}\ \bibnamefont
  {Okamura}}, \bibinfo {author} {\bibfnamefont {Yastoshi}\ \bibnamefont
  {Takayama}}, \ and\ \bibinfo {author} {\bibfnamefont {Kentaroh}\ \bibnamefont
  {Yoshida}},\ }\bibfield  {title} {\enquote {\bibinfo {title} {{Open spinning
  strings and AdS/dCFT duality}},}\ }\href {\doibase
  10.1088/1126-6708/2006/01/112} {\bibfield  {journal} {\bibinfo  {journal}
  {JHEP}\ }\textbf {\bibinfo {volume} {01}},\ \bibinfo {pages} {112} (\bibinfo
  {year} {2006})},\ \Eprint {http://arxiv.org/abs/hep-th/0511139}
  {arXiv:hep-th/0511139 [hep-th]} \BibitemShut {NoStop}%
\bibitem [{\citenamefont {Nagasaki}\ and\ \citenamefont
  {Yamaguchi}(2012)}]{Nagasaki:2012re}%
  \BibitemOpen
  \bibfield  {author} {\bibinfo {author} {\bibfnamefont {Koichi}\ \bibnamefont
  {Nagasaki}}\ and\ \bibinfo {author} {\bibfnamefont {Satoshi}\ \bibnamefont
  {Yamaguchi}},\ }\bibfield  {title} {\enquote {\bibinfo {title} {{Expectation
  values of chiral primary operators in holographic interface CFT}},}\ }\href
  {\doibase 10.1103/PhysRevD.86.086004} {\bibfield  {journal} {\bibinfo
  {journal} {Phys. Rev.}\ }\textbf {\bibinfo {volume} {D86}},\ \bibinfo {pages}
  {086004} (\bibinfo {year} {2012})},\ \Eprint {http://arxiv.org/abs/1205.1674}
  {arXiv:1205.1674 [hep-th]} \BibitemShut {NoStop}%
\bibitem [{\citenamefont {Kristjansen}\ \emph {et~al.}(2013)\citenamefont
  {Kristjansen}, \citenamefont {Semenoff},\ and\ \citenamefont
  {Young}}]{Kristjansen:2012tn}%
  \BibitemOpen
  \bibfield  {author} {\bibinfo {author} {\bibfnamefont {Charlotte}\
  \bibnamefont {Kristjansen}}, \bibinfo {author} {\bibfnamefont {Gordon~W.}\
  \bibnamefont {Semenoff}}, \ and\ \bibinfo {author} {\bibfnamefont {Donovan}\
  \bibnamefont {Young}},\ }\bibfield  {title} {\enquote {\bibinfo {title}
  {{Chiral primary one-point functions in the D3-D7 defect conformal field
  theory}},}\ }\href {\doibase 10.1007/JHEP01(2013)117} {\bibfield  {journal}
  {\bibinfo  {journal} {JHEP}\ }\textbf {\bibinfo {volume} {01}},\ \bibinfo
  {pages} {117} (\bibinfo {year} {2013})},\ \Eprint
  {http://arxiv.org/abs/1210.7015} {arXiv:1210.7015 [hep-th]} \BibitemShut
  {NoStop}%
\bibitem [{\citenamefont {Minahan}\ and\ \citenamefont
  {Zarembo}(2003)}]{Minahan:2002ve}%
  \BibitemOpen
  \bibfield  {author} {\bibinfo {author} {\bibfnamefont {J.~A.}\ \bibnamefont
  {Minahan}}\ and\ \bibinfo {author} {\bibfnamefont {K.}~\bibnamefont
  {Zarembo}},\ }\bibfield  {title} {\enquote {\bibinfo {title} {{The Bethe
  ansatz for $\mathcal{N}=4$ superYang-Mills}},}\ }\href {\doibase
  10.1088/1126-6708/2003/03/013} {\bibfield  {journal} {\bibinfo  {journal}
  {JHEP}\ }\textbf {\bibinfo {volume} {03}},\ \bibinfo {pages} {013} (\bibinfo
  {year} {2003})},\ \Eprint {http://arxiv.org/abs/hep-th/0212208}
  {arXiv:hep-th/0212208 [hep-th]} \BibitemShut {NoStop}%
\bibitem [{\citenamefont {Beisert}\ and\ \citenamefont
  {Staudacher}(2003)}]{Beisert:2003yb}%
  \BibitemOpen
  \bibfield  {author} {\bibinfo {author} {\bibfnamefont {Niklas}\ \bibnamefont
  {Beisert}}\ and\ \bibinfo {author} {\bibfnamefont {Matthias}\ \bibnamefont
  {Staudacher}},\ }\bibfield  {title} {\enquote {\bibinfo {title} {{The
  $\mathcal{N}=4$ SYM integrable super spin chain}},}\ }\href {\doibase
  10.1016/j.nuclphysb.2003.08.015} {\bibfield  {journal} {\bibinfo  {journal}
  {Nucl. Phys.}\ }\textbf {\bibinfo {volume} {B670}},\ \bibinfo {pages}
  {439--463} (\bibinfo {year} {2003})},\ \Eprint
  {http://arxiv.org/abs/hep-th/0307042} {arXiv:hep-th/0307042 [hep-th]}
  \BibitemShut {NoStop}%
\bibitem [{\citenamefont {de~Leeuw}\ \emph {et~al.}(2015)\citenamefont
  {de~Leeuw}, \citenamefont {Kristjansen},\ and\ \citenamefont
  {Zarembo}}]{deLeeuw:2015hxa}%
  \BibitemOpen
  \bibfield  {author} {\bibinfo {author} {\bibfnamefont {Marius}\ \bibnamefont
  {de~Leeuw}}, \bibinfo {author} {\bibfnamefont {Charlotte}\ \bibnamefont
  {Kristjansen}}, \ and\ \bibinfo {author} {\bibfnamefont {Konstantin}\
  \bibnamefont {Zarembo}},\ }\bibfield  {title} {\enquote {\bibinfo {title}
  {{One-point Functions in Defect CFT and Integrability}},}\ }\href {\doibase
  10.1007/JHEP08(2015)098} {\bibfield  {journal} {\bibinfo  {journal} {JHEP}\
  }\textbf {\bibinfo {volume} {08}},\ \bibinfo {pages} {098} (\bibinfo {year}
  {2015})},\ \Eprint {http://arxiv.org/abs/1506.06958} {arXiv:1506.06958
  [hep-th]} \BibitemShut {NoStop}%
\bibitem [{\citenamefont {Buhl-Mortensen}\ \emph {et~al.}(2016)\citenamefont
  {Buhl-Mortensen}, \citenamefont {de~Leeuw}, \citenamefont {Kristjansen},\
  and\ \citenamefont {Zarembo}}]{Buhl-Mortensen:2015gfd}%
  \BibitemOpen
  \bibfield  {author} {\bibinfo {author} {\bibfnamefont {Isak}\ \bibnamefont
  {Buhl-Mortensen}}, \bibinfo {author} {\bibfnamefont {Marius}\ \bibnamefont
  {de~Leeuw}}, \bibinfo {author} {\bibfnamefont {Charlotte}\ \bibnamefont
  {Kristjansen}}, \ and\ \bibinfo {author} {\bibfnamefont {Konstantin}\
  \bibnamefont {Zarembo}},\ }\bibfield  {title} {\enquote {\bibinfo {title}
  {{One-point Functions in AdS/dCFT from Matrix Product States}},}\ }\href
  {\doibase 10.1007/JHEP02(2016)052} {\bibfield  {journal} {\bibinfo  {journal}
  {JHEP}\ }\textbf {\bibinfo {volume} {02}},\ \bibinfo {pages} {052} (\bibinfo
  {year} {2016})},\ \Eprint {http://arxiv.org/abs/1512.02532} {arXiv:1512.02532
  [hep-th]} \BibitemShut {NoStop}%
\bibitem [{\citenamefont {de~Leeuw}\ \emph {et~al.}(2016)\citenamefont
  {de~Leeuw}, \citenamefont {Kristjansen},\ and\ \citenamefont
  {Mori}}]{deLeeuw:2016umh}%
  \BibitemOpen
  \bibfield  {author} {\bibinfo {author} {\bibfnamefont {Marius}\ \bibnamefont
  {de~Leeuw}}, \bibinfo {author} {\bibfnamefont {Charlotte}\ \bibnamefont
  {Kristjansen}}, \ and\ \bibinfo {author} {\bibfnamefont {Stefano}\
  \bibnamefont {Mori}},\ }\bibfield  {title} {\enquote {\bibinfo {title}
  {{AdS/dCFT one-point functions of the SU(3) sector}},}\ }\href {\doibase
  10.1016/j.physletb.2016.10.044} {\bibfield  {journal} {\bibinfo  {journal}
  {Phys. Lett.}\ }\textbf {\bibinfo {volume} {B763}},\ \bibinfo {pages}
  {197--202} (\bibinfo {year} {2016})},\ \Eprint
  {http://arxiv.org/abs/1607.03123} {arXiv:1607.03123 [hep-th]} \BibitemShut
  {NoStop}%
\bibitem [{\citenamefont {Buhl-Mortensen}\ \emph {et~al.}()\citenamefont
  {Buhl-Mortensen}, \citenamefont {de~Leeuw}, \citenamefont {Ipsen},
  \citenamefont {Kristjansen},\ and\ \citenamefont {Wilhelm}}]{longpaper}%
  \BibitemOpen
  \bibfield  {author} {\bibinfo {author} {\bibfnamefont {Isak}\ \bibnamefont
  {Buhl-Mortensen}}, \bibinfo {author} {\bibfnamefont {Marius}\ \bibnamefont
  {de~Leeuw}}, \bibinfo {author} {\bibfnamefont {Asger~C.}\ \bibnamefont
  {Ipsen}}, \bibinfo {author} {\bibfnamefont {Charlotte}\ \bibnamefont
  {Kristjansen}}, \ and\ \bibinfo {author} {\bibfnamefont {Matthias}\
  \bibnamefont {Wilhelm}},\ }\bibfield  {title} {\enquote {\bibinfo {title} {{A
  Quantum Check of AdS/dCFT}},}\ }\href@noop {} {\ }\bibinfo {note}
  {\!\!\scalebox{1.15}{\!}\scalebox{1.2}{\textcolor{white}{.}}\,to
  appear}\BibitemShut {NoStop}%
\bibitem [{\citenamefont {Erdmenger}\ \emph {et~al.}(2002)\citenamefont
  {Erdmenger}, \citenamefont {Guralnik},\ and\ \citenamefont
  {Kirsch}}]{Erdmenger:2002ex}%
  \BibitemOpen
  \bibfield  {author} {\bibinfo {author} {\bibfnamefont {Johanna}\ \bibnamefont
  {Erdmenger}}, \bibinfo {author} {\bibfnamefont {Zachary}\ \bibnamefont
  {Guralnik}}, \ and\ \bibinfo {author} {\bibfnamefont {Ingo}\ \bibnamefont
  {Kirsch}},\ }\bibfield  {title} {\enquote {\bibinfo {title}
  {{Four-dimensional superconformal theories with interacting boundaries or
  defects}},}\ }\href {\doibase 10.1103/PhysRevD.66.025020} {\bibfield
  {journal} {\bibinfo  {journal} {Phys. Rev.}\ }\textbf {\bibinfo {volume}
  {D66}},\ \bibinfo {pages} {025020} (\bibinfo {year} {2002})},\ \Eprint
  {http://arxiv.org/abs/hep-th/0203020} {arXiv:hep-th/0203020 [hep-th]}
  \BibitemShut {NoStop}%
\bibitem [{\citenamefont {Breitenlohner}\ and\ \citenamefont
  {Freedman}(1982)}]{Breitenlohner:1982jf}%
  \BibitemOpen
  \bibfield  {author} {\bibinfo {author} {\bibfnamefont {Peter}\ \bibnamefont
  {Breitenlohner}}\ and\ \bibinfo {author} {\bibfnamefont {Daniel~Z.}\
  \bibnamefont {Freedman}},\ }\bibfield  {title} {\enquote {\bibinfo {title}
  {{Stability in Gauged Extended Supergravity}},}\ }\href {\doibase
  10.1016/0003-4916(82)90116-6} {\bibfield  {journal} {\bibinfo  {journal}
  {Annals Phys.}\ }\textbf {\bibinfo {volume} {144}},\ \bibinfo {pages} {249}
  (\bibinfo {year} {1982})}\BibitemShut {NoStop}%
\bibitem [{\citenamefont {Allen}\ and\ \citenamefont
  {Jacobson}(1986)}]{Allen:1985wd}%
  \BibitemOpen
  \bibfield  {author} {\bibinfo {author} {\bibfnamefont {Bruce}\ \bibnamefont
  {Allen}}\ and\ \bibinfo {author} {\bibfnamefont {Theodore}\ \bibnamefont
  {Jacobson}},\ }\bibfield  {title} {\enquote {\bibinfo {title} {{Vector Two
  Point Functions in Maximally Symmetric Spaces}},}\ }\href {\doibase
  10.1007/BF01211169} {\bibfield  {journal} {\bibinfo  {journal} {Commun. Math.
  Phys.}\ }\textbf {\bibinfo {volume} {103}},\ \bibinfo {pages} {669} (\bibinfo
  {year} {1986})}\BibitemShut {NoStop}%
\bibitem [{\citenamefont {Camporesi}(1991)}]{Camporesi91}%
  \BibitemOpen
  \bibfield  {author} {\bibinfo {author} {\bibfnamefont {Roberto}\ \bibnamefont
  {Camporesi}},\ }\bibfield  {title} {\enquote {\bibinfo {title}
  {$\ensuremath{\zeta}$-function regularization of one-loop effective
  potentials in anti-de sitter spacetime},}\ }\href {\doibase
  10.1103/PhysRevD.43.3958} {\bibfield  {journal} {\bibinfo  {journal} {Phys.
  Rev. D}\ }\textbf {\bibinfo {volume} {43}},\ \bibinfo {pages} {3958--3965}
  (\bibinfo {year} {1991})}\BibitemShut {NoStop}%
\bibitem [{\citenamefont {Liu}\ and\ \citenamefont
  {Tseytlin}(1999)}]{Liu:1998ty}%
  \BibitemOpen
  \bibfield  {author} {\bibinfo {author} {\bibfnamefont {Hong}\ \bibnamefont
  {Liu}}\ and\ \bibinfo {author} {\bibfnamefont {Arkady~A.}\ \bibnamefont
  {Tseytlin}},\ }\bibfield  {title} {\enquote {\bibinfo {title} {{On four point
  functions in the CFT / AdS correspondence}},}\ }\href {\doibase
  10.1103/PhysRevD.59.086002} {\bibfield  {journal} {\bibinfo  {journal} {Phys.
  Rev.}\ }\textbf {\bibinfo {volume} {D59}},\ \bibinfo {pages} {086002}
  (\bibinfo {year} {1999})},\ \Eprint {http://arxiv.org/abs/hep-th/9807097}
  {arXiv:hep-th/9807097 [hep-th]} \BibitemShut {NoStop}%
\bibitem [{\citenamefont {Nagasaki}\ \emph {et~al.}(2012)\citenamefont
  {Nagasaki}, \citenamefont {Tanida},\ and\ \citenamefont
  {Yamaguchi}}]{Nagasaki:2011ue}%
  \BibitemOpen
  \bibfield  {author} {\bibinfo {author} {\bibfnamefont {Koichi}\ \bibnamefont
  {Nagasaki}}, \bibinfo {author} {\bibfnamefont {Hiroaki}\ \bibnamefont
  {Tanida}}, \ and\ \bibinfo {author} {\bibfnamefont {Satoshi}\ \bibnamefont
  {Yamaguchi}},\ }\bibfield  {title} {\enquote {\bibinfo {title} {{Holographic
  Interface-Particle Potential}},}\ }\href {\doibase 10.1007/JHEP01(2012)139}
  {\bibfield  {journal} {\bibinfo  {journal} {JHEP}\ }\textbf {\bibinfo
  {volume} {01}},\ \bibinfo {pages} {139} (\bibinfo {year} {2012})},\ \Eprint
  {http://arxiv.org/abs/1109.1927} {arXiv:1109.1927 [hep-th]} \BibitemShut
  {NoStop}%
\bibitem [{Note1()}]{Note1}%
  \BibitemOpen
  \bibinfo {note} {It can be found e.g.\ in~\cite
  {deLeeuw:2015hxa}}\BibitemShut {NoStop}%
\bibitem [{\citenamefont {Siegel}(1979)}]{Siegel:1979wq}%
  \BibitemOpen
  \bibfield  {author} {\bibinfo {author} {\bibfnamefont {Warren}\ \bibnamefont
  {Siegel}},\ }\bibfield  {title} {\enquote {\bibinfo {title} {{Supersymmetric
  Dimensional Regularization via Dimensional Reduction}},}\ }\href {\doibase
  10.1016/0370-2693(79)90282-X} {\bibfield  {journal} {\bibinfo  {journal}
  {Phys. Lett.}\ }\textbf {\bibinfo {volume} {B84}},\ \bibinfo {pages}
  {193--196} (\bibinfo {year} {1979})}\BibitemShut {NoStop}%
\bibitem [{\citenamefont {Witten}(1998)}]{Witten:1998qj}%
  \BibitemOpen
  \bibfield  {author} {\bibinfo {author} {\bibfnamefont {Edward}\ \bibnamefont
  {Witten}},\ }\bibfield  {title} {\enquote {\bibinfo {title} {{Anti-de Sitter
  space and holography}},}\ }\href@noop {} {\bibfield  {journal} {\bibinfo
  {journal} {Adv. Theor. Math. Phys.}\ }\textbf {\bibinfo {volume} {2}},\
  \bibinfo {pages} {253--291} (\bibinfo {year} {1998})},\ \Eprint
  {http://arxiv.org/abs/hep-th/9802150} {arXiv:hep-th/9802150 [hep-th]}
  \BibitemShut {NoStop}%
\bibitem [{\citenamefont {Berenstein}\ \emph {et~al.}(2002)\citenamefont
  {Berenstein}, \citenamefont {Maldacena},\ and\ \citenamefont
  {Nastase}}]{Berenstein:2002jq}%
  \BibitemOpen
  \bibfield  {author} {\bibinfo {author} {\bibfnamefont {David~Eliecer}\
  \bibnamefont {Berenstein}}, \bibinfo {author} {\bibfnamefont {Juan~Martin}\
  \bibnamefont {Maldacena}}, \ and\ \bibinfo {author} {\bibfnamefont
  {Horatiu~Stefan}\ \bibnamefont {Nastase}},\ }\bibfield  {title} {\enquote
  {\bibinfo {title} {{Strings in flat space and pp waves from $\mathcal{N}=4$
  superYang-Mills}},}\ }\href {\doibase 10.1088/1126-6708/2002/04/013}
  {\bibfield  {journal} {\bibinfo  {journal} {JHEP}\ }\textbf {\bibinfo
  {volume} {04}},\ \bibinfo {pages} {013} (\bibinfo {year} {2002})},\ \Eprint
  {http://arxiv.org/abs/hep-th/0202021} {arXiv:hep-th/0202021 [hep-th]}
  \BibitemShut {NoStop}%
\bibitem [{\citenamefont {Beisert}\ \emph {et~al.}(2007)\citenamefont
  {Beisert}, \citenamefont {Eden},\ and\ \citenamefont
  {Staudacher}}]{Beisert:2006ez}%
  \BibitemOpen
  \bibfield  {author} {\bibinfo {author} {\bibfnamefont {Niklas}\ \bibnamefont
  {Beisert}}, \bibinfo {author} {\bibfnamefont {Burkhard}\ \bibnamefont
  {Eden}}, \ and\ \bibinfo {author} {\bibfnamefont {Matthias}\ \bibnamefont
  {Staudacher}},\ }\bibfield  {title} {\enquote {\bibinfo {title}
  {{Transcendentality and Crossing}},}\ }\href {\doibase
  10.1088/1742-5468/2007/01/P01021} {\bibfield  {journal} {\bibinfo  {journal}
  {J. Stat. Mech.}\ }\textbf {\bibinfo {volume} {0701}},\ \bibinfo {pages}
  {P01021} (\bibinfo {year} {2007})},\ \Eprint
  {http://arxiv.org/abs/hep-th/0610251} {arXiv:hep-th/0610251 [hep-th]}
  \BibitemShut {NoStop}%
\bibitem [{\citenamefont {Bern}\ \emph {et~al.}(2007)\citenamefont {Bern},
  \citenamefont {Czakon}, \citenamefont {Dixon}, \citenamefont {Kosower},\ and\
  \citenamefont {Smirnov}}]{Bern:2006ew}%
  \BibitemOpen
  \bibfield  {author} {\bibinfo {author} {\bibfnamefont {Zvi}\ \bibnamefont
  {Bern}}, \bibinfo {author} {\bibfnamefont {Michael}\ \bibnamefont {Czakon}},
  \bibinfo {author} {\bibfnamefont {Lance~J.}\ \bibnamefont {Dixon}}, \bibinfo
  {author} {\bibfnamefont {David~A.}\ \bibnamefont {Kosower}}, \ and\ \bibinfo
  {author} {\bibfnamefont {Vladimir~A.}\ \bibnamefont {Smirnov}},\ }\bibfield
  {title} {\enquote {\bibinfo {title} {{The Four-Loop Planar Amplitude and Cusp
  Anomalous Dimension in Maximally Supersymmetric Yang-Mills Theory}},}\ }\href
  {\doibase 10.1103/PhysRevD.75.085010} {\bibfield  {journal} {\bibinfo
  {journal} {Phys. Rev.}\ }\textbf {\bibinfo {volume} {D75}},\ \bibinfo {pages}
  {085010} (\bibinfo {year} {2007})},\ \Eprint
  {http://arxiv.org/abs/hep-th/0610248} {arXiv:hep-th/0610248 [hep-th]}
  \BibitemShut {NoStop}%
\bibitem [{\citenamefont {Cachazo}\ \emph {et~al.}(2007)\citenamefont
  {Cachazo}, \citenamefont {Spradlin},\ and\ \citenamefont
  {Volovich}}]{Cachazo:2006az}%
  \BibitemOpen
  \bibfield  {author} {\bibinfo {author} {\bibfnamefont {Freddy}\ \bibnamefont
  {Cachazo}}, \bibinfo {author} {\bibfnamefont {Marcus}\ \bibnamefont
  {Spradlin}}, \ and\ \bibinfo {author} {\bibfnamefont {Anastasia}\
  \bibnamefont {Volovich}},\ }\bibfield  {title} {\enquote {\bibinfo {title}
  {{Four-loop cusp anomalous dimension from obstructions}},}\ }\href {\doibase
  10.1103/PhysRevD.75.105011} {\bibfield  {journal} {\bibinfo  {journal} {Phys.
  Rev.}\ }\textbf {\bibinfo {volume} {D75}},\ \bibinfo {pages} {105011}
  (\bibinfo {year} {2007})},\ \Eprint {http://arxiv.org/abs/hep-th/0612309}
  {arXiv:hep-th/0612309 [hep-th]} \BibitemShut {NoStop}%
\end{thebibliography}%

\end{document}